\RequirePackage{fix-cm}
\documentclass[smallextended]{svjour3}       
\smartqed  
\usepackage{graphicx}

\usepackage{amsmath,amssymb,amsfonts,mathtools}
\usepackage{epsfig}
\usepackage{dcolumn}
\usepackage{tabularx}
\usepackage[dvipsnames]{color}

\usepackage[normalem]{ulem} 
\usepackage{soul} %

\begin{document}

\title{Improvements on Particle Tracking Velocimetry: model-free calibration and noiseless measurement of second order statistics of the velocity field}
\titlerunning{Improvements on Particle Tracking Velocimetry}

\author{Nathana\"el Machicoane \and Miguel L\'opez-Caballero \and Mickael Bourgoin \and  Alberto Aliseda \and Romain Volk}

\authorrunning{N. Machicoane \and M. L\'opez-Caballero \and M. Bourgoin \and  A. Aliseda \and R. Volk} 

\institute{Alberto Aliseda \at
              University of Washington - Department of Mechanical Engineering, Seattle, WA, USA 
           \and
           Mickael Bourgoin \at
              LEGI, Universit\' de Grenoble/G-INP/UJF/CRNS, BP53 Grenoble, 38041 cedex 9, France\\
              Univ Lyon, Ens de Lyon, Univ Claude Bernard, CNRS, Laboratoire de Physique, F-69342 Lyon, France              
              \and
              Nathana\"el Machicoane \at Laboratoire FAST, CNRS, Universit\'e Paris-Sud,
Orsay, France              
              \and 
              Miguel L\'opez-Caballero \at Univ Lyon, Ens de Lyon, Univ Claude Bernard, CNRS, Laboratoire de Physique, F-69342 Lyon, France
              \and
              Romain Volk \at Univ Lyon, Ens de Lyon, Univ Claude Bernard, CNRS, Laboratoire de Physique, F-69342 Lyon, France \\
              Tel.: +33-4-72-72-39-45\\
              Fax: +33-4-72-72-89-50\\
              \email{romain.volk@ens-lyon.fr}  
}

\date{Received: date / Accepted: date}

\maketitle

\begin{abstract}
This article describes two independent developments aimed at improving the Particle Tracking Method for measurements of flow or particle velocities. First, a stereoscopic multicamera calibration method that does not require any optical model is described and evaluated. We show that this new calibration method gives better results than the most commonly-used technique, based on the Tsai camera/optics model. Additionally, the methods uses a simple interpolant to compute the transformation matrix and it is trivial to apply for any experimental fluid dynamics visualization set up. The second contribution proposes a solution to remove noise from Eulerian  measurements of velocity statistics obtained from Particle Tracking velocimetry, without the need of filtering and/or windowing. The novel method presented here is based on recomputing particle displacement measurements from two consecutive frames for multiple different time-step values between frames. We show the successful application of this new technique to recover the second order velocity structure function of the flow. Increased accuracy is demonstrated by comparing the dissipation rate of turbulent kinetic energy measured from the second order structure function against previously validated measurements. These two techniques for improvement of experimental fluid/particle velocity measurements can be combined to provide high accuracy 3D particle and/or flow velocity statistics and derived variables needed to characterize a turbulent flow.
\end{abstract}

\section{Introduction}
Flow velocity measurements, based on the analysis of the motion of particles imaged with digital cameras, have become the most commonly-used metrology technique in contemporary fluid mechanics research~\cite{adrian1991particle}. \emph{Particle Image Velocimetry} (PIV) and \emph{Particle Tracking Velocimetry} (PTV) are two widely used methods that enable the characterization of a flow from the Eulerian (PIV) or Lagrangian (PTV) point of view. Several aspects influence the accuracy and reliability of the measurements obtained with these techniques: resolution (temporal and spatial), dynamical range (spatial and temporal), capacity to measure 2D or 3D components of velocity in a 2D or 3D fluid domain, statistical convergence, etc... These imaging and analysis considerations depend on the hardware (camera's resolution, repetition rate, on board memory, optical system, etc.) but also on the software (optical calibration relating real world coordinates to pixel coordinates, particle identification and tracking algorithms, image correlations, dynamical post-processing, etc.) used in the measurements.

The particle tracking velocimetry method is a widely used experimental techniques that can provide highly resolved, in space and time, measurements of the flow velocity (if the particles are flow tracers) or particulate velocities (if the particles are not just following the underlying flow velocity) in experimental fluid mechanics and applications {\cite{bib:sato,bib:virant,Voth:jfm2002,bib:ouelette}}. Recently, stereoscopic velocity measurements of the three components of the velocity in a relatively thin volume (3C-2D) and fully three dimensional measurements of the three components of the velocity at a prismatic volume in the flow (3C-3D) using multi camera approaches and wider illumination regions have become widely available for particle tracking, as for PIV. Two frequent realizations of this method in the laboratory are based on taking {a pair of} images (with double exposure cameras, typical of PIV) at very short time separation followed by a large{r} time interval, and collecting a long sequence of images closely and equally separated in time (with high speed cameras). In the first case, the Particle Tracking Velocimetry technique provides a single vector per particle in a pair of consecutive images, with subsequent velocity measurements in other image pairs being uncorrelated from this. The high speed image sequence, on the contrary, provides the opportunity to track the same particle over multiple (N) images and provides several (N-1) correlated velocity measurements, at different locations but along the same particle trajectory. There are two contributions in this paper that apply equally to both versions of the Particle Tracking Velocimetry technique: each one advances the state of the art in a stage of the measurement of velocity from particle images. The first contribution is to provide an optical-model-free calibration technique for multi camera particle tracking velocimetry and potentially also for Particle Image Velocimetry. This method is simpler to apply, provides equal or better results that the commonly-used Tsai model calibration, and is computationally efficient to apply on the images in {a classical} PTV algorithm workflow {for instance}. The second contribution is to provide a method to remove noise from Particle Tracking Velocimetry in the calculation of statistics moments of the velocities from the PTV measurements. This is a common operation in fluid mechanics, for example in computing Eulerian velocity statistics by binning in space the PTV measurements, to determine flow features and turbulence characteristics (such as dissipation, ...). The technique relies on multiple sets of images collected at different time separations, which can be achieved with buffered double-exposure cameras (\`a la PIV) by repeating the experiments while setting the $\Delta t$ between laser pulses and camera exposures to different values or with a sequence of high speed images by skipping a varying number {of} images between a pair in the tracking algorithm. The result is an estimate of the velocity statistics without noise: the noise is subtracted since it is not correlated to $\Delta t$. This may or may not be the case in traditional PIV at different time separations, depending on the flow and settings. Thus, it is possible to explore the application of this technique to flow fields computed with PIV at different values of $\Delta t$. 

Although they are strongest if applied in combination, the two contributions are described in the following sections in an independent manner. To facilitate their application, first the calibration method and a comparison of its results the Tsai model in a canonical flow are presented. Then, the noiseless velocity statistics computation is shown with an example in homogeneous isotropic turbulence where independent measurements were already available and the effect of the noise elimination is shown in the results and comparison with the benchmark measurements collected via other non optical methods.


\section{Novel Method for Optical Calibration of a multi-camera 3D Particle Imaging or Tracking Velocimetry System}\label{sec:calibration}

\subsection{On stereoscopic camera calibration}
Particle tracking velocimetry involves the visualization, with at least two cameras for 3D measurements, of a fluid domain (the measurement volume) via light scattered by small tracer particles with which the flow has been seeded. For PTV, individual particles are identified in successive frames and associated as images of the same particle to reconstruct the particle trajectory over time. PTV is, therefore, an intrinsically Lagrangian method (though, as it will be discussed later, such Lagrangian data can be used to compute Eulerian statistics. Thanks to the impressive improvements of high speed imaging technologies during the last decade (contemporary cameras can record mega-pixel images at rate of tens of thousands frames per second), this method has become one of the most accurate velocity measurement techniques in fluid mechanics. Specifically, it has become a standard measurement technique for the investigation of turbulence, a research field that is particularly demanding measurement-wise due to its intrinsic multi-scale and three dimensional nature. 

The 3D trajectory reconstruction is usually done in several steps: 
\begin{enumerate}
\item particle centers are identified by image processing, in pixel coordinates relative to each camera, for each recorded frame (at each time step and for each camera), 
\item the particle pixel coordinates from each camera are converted to a vector of possible positions in physical space (using a calibration scheme). The most probable location for the intersection of all the particle vectors (one coming from each camera) is computed by the stereo-matching procedure, thus determining the 3D position of each particle in real world coordinates, at each time step,
\item the 3D positions of a particle at different time steps (consecutive frames) are connected to reconstruct the particle trajectory.

\end{enumerate}

\begin{figure}
\centering
\includegraphics[width=.7\columnwidth]{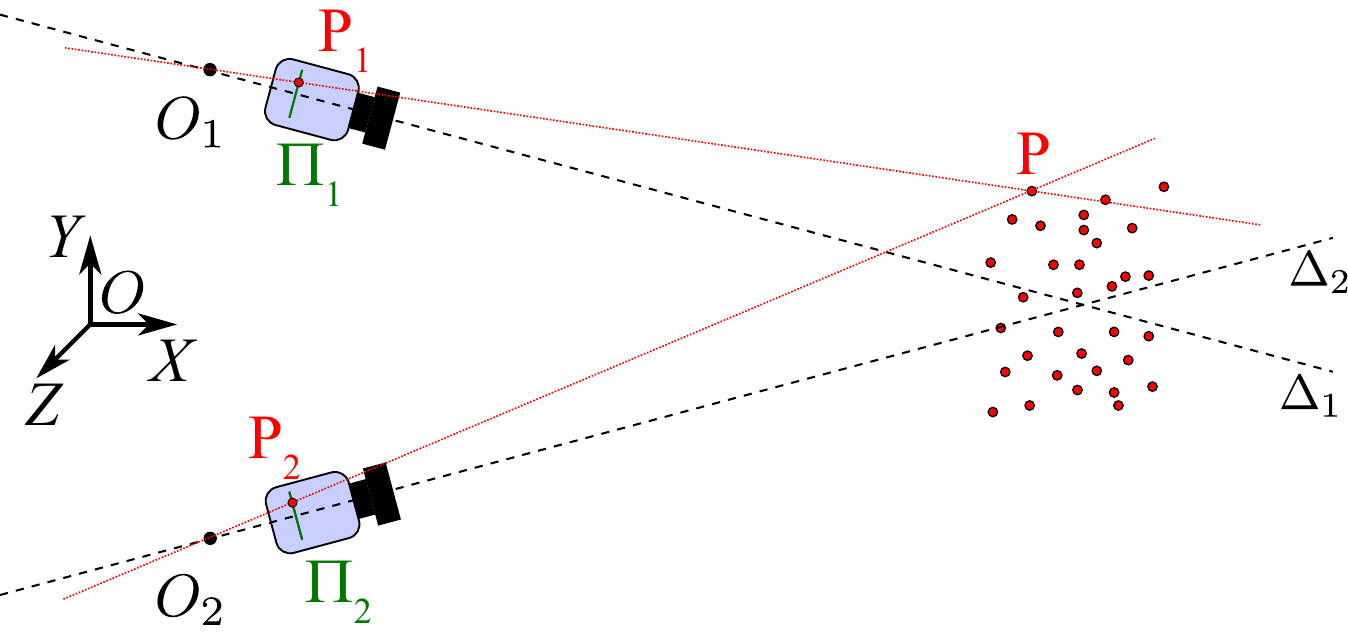}
\caption{Pinhole Tsai's camera model and 3D stereo-matching: the 3D position of a particle corresponds to the intersection of 2 lines $\Delta_1$ and $\Delta_2$, each emitted by the camera center $O_1$ and $O_2$ and passing through the {2D position of a detected particle $P_1$ and $P_2$ on a each camera plane $\Pi_1$ and $\Pi_2$.}}\label{fig:Tsai}
\end{figure}

Steps 1 and 3 have been extensively studied in image processing and particle kinematics analysis in the literature. For example, Ouellette and collaborators~\cite{bib:ouellette2005} have compared the robustness and accuracy of several particle detection (center of mass, 2D Gaussian fittings, 2x1D Gaussian fitting, neural networks algorithms, etc.) and trajectory reconstruction methods (nearest neighbor, 2 times prediction-correction methods, 3 times prediction-correction methods, etc.). Step 2, optical calibration and 3D position determination, has been the subject of much less attention. Almost all existing experimental implementations of multi-camera Particle Tracking use the pinhole camera model, originally proposed by Tsai in 1987~\cite{bib:tsai1987} as the basis for calibration. In Tsai's approach, each camera ${\cal{C}}_i$ is replaced by a pinhole, defined by: (i) one pinhole center $O_i$ (with coordinates $(X_{0_i}, Y_{0_i}, Z_{0_i})$), (ii) one optical axis $\Delta_i$ (passing by $O_i$, with angular orientations $(\alpha_{\Delta_i}, \beta_{\Delta_i}, \gamma_{\Delta_i})$), (iii) one projection plane $P_i$, perpendicular to $\Delta_i$ and at a distance $f_i$ from $O_i$ ($f_i$ is called the \emph{focal length} of the pinhole model). Each camera is therefore modeled with at least 7 parameters (others can be included in the model to account for  optical aberrations, aspect ratio of individual pixels, etc.). In this model, the image $P_i$ of a particle $P$ on camera ${\cal{C}}_i$ is by definition the intersection of the line $(PO_i)$ with the plane $\Pi_i$. Conversely, given a particle image $P_i$ on the plane $\Pi_i$, we know that the actual particle lies somewhere along the vector $(O_i P_i)$. 3D position is then simply obtained from the stereo-matching of the images of a particle, calculating the most probable intersection point between the vectors, $(O_1 P_1)\cap (O_2P_2)$, on at least two cameras, ${\cal{C}}_1$ and ${\cal{C}}_2$ (Fig.~\ref{fig:Tsai}).

Calibration approaches based on Tsai model relies on a very simple linear model (with eventual non-linear corrections to account for optical aberrations) which is intuitive and requires a relatively minimal number of parameters to describe each camera. These can represent a significant advantage in setting up the calibration algorithm, but this simplicity also implies important drawbacks:
\begin{itemize}
\item although nonlinear optical aberration can be implemented in the model, it is difficult in such a simple description to account for strong optical deformations due for instance to imperfections of diopter surfaces.
\item Tsai's model lacks versatility when non-standard optics are used, for instance when Scheimpflug mounts are required, or when the imaging is done through a diopter at an angle (oblique observation through a flat window in a water tank, for instance). In these cases, further additions to the model need to be implemented~\cite{bib:louhichi}, introducing  additional parameter models that must be determined a priori.
\item The line of sight reconstructed by the model is an approximation of the actual path followed by light from the scattering particle to the imaging sensor. As the model assumes a straight path, it cannot handle stratified flows with a gradient of refractive index, or setups with multiple refractive interfaces.
\end{itemize}

A new camera calibration method is described here that can be up to 5 times more accurate than Tsai's model in terms of absolute 3D stereo-positioning of particles, while easily handling any of the complexity or non-linearity in the optical setup described above. The key point of the new  method is that, instead of a priori defining an optical model of the imaging system, it defines an interpolant that connects each point in the camera sensor to the actual light beam across the measurement volume. This interpolant contains {the necessary} degrees of freedom but does not require any operator input beyond a set of calibration images taken across the measurement volume (typical of any 3D calibration process).  The description of the method and its quantitative characterization compared to the traditional Tsai model are given in {the following sub-sections}.

\subsection{Calibration principle}
3D particle imaging methods require an appropriate calibration method to perform the correspondence and the stereo-matching between the several sets of 2D positions of particles in the \emph{pixels coordinate system} for each camera and the absolute 3D position of the particles in \emph{real world coordinate system}. The accuracy of the calibration method directly impacts the accuracy of the absolute 3D positioning of the particles in real world. We will not discuss in this article methods for center finding of particles in pixel units in the 2D pixel images. A deep discussion of this aspect can be found in the literature, for instance in~\cite{bib:ouellette2005}. 

The new calibration method we propose here for the stereo-matching step is based on a very simple idea: no matter how distorted an image recorded on the pixel array of a camera can be, each bright point of the pixel array can be associated to a ray of light which produced it, such that the corresponding light source (typically a scatterer particle) can lie anywhere on this ray of light. An appropriate calibration method should therefore be able to directly attribute to a given doublet $(x_p,y_p)$ of pixel coordinates the corresponding ray path. If the index of refraction in the measurement volume of interest is homogeneous (so that light propagates along a straight line) each doublet $(x_p,y_p)$ can therefore be associated to a straight line $\Delta$ (defined by 6 parameters in 3D: one point $O_\Delta(x_p,y_p)$ and one vector $\vec{V}_\Delta(x_p,y_p)$), regardless of the path outside the volume of interest, which can be very complex as interfaces and lenses can be crossed. The method we propose consists simply in building a \emph{pixel to line} interpolant $\cal I$ to perform this correspondence between pixel coordinates and each of the 6 parameters of the ray of light:

\begin{equation}
(x_p,y_p) \xrightarrow{{\cal{I}}} (O_\Delta,\vec{V}_\Delta)
\end{equation}

This can seem similar at first sight to the Tsai approach, which for a given $(x_p,y_p)$ doublet also builds a ray of light. The big difference here is that while the Tsai approach assumes a model for the camera (namely a pinhole model), and is therefore sensitive to imperfections of the model, our approach does not relie on any \emph{a priori} model and is only based on empirical interpolations adapted to the actual calibration data. It can therefore self-adapt to optical imperfections, media inhomogeneities or refined cameras arrangements. For instance, the generalization of the method to cases where light does not propagate in a straight line (as in stratified fluids for instance) is straightforward as it is sufficient to build the interpolant not for the parameters defining a line, but for the parameters required to describe the expected curve for the path of light in the media (for instance a parabola in a linearly stratified medium). Besides, since the pixel-line correspondence does not rely on any \emph{a priori} camera model, it is very robust and its accuracy only depends on the precision of the built interpolant. The following sub-section describes the practical implementation of this idea, and a simple protocol to accurately build the interpolant ${\cal{I}}$.

\subsection{Practical Implementation}
We propose in this section a simple implementation of the previous idea in order to build the interpolant $\cal{I}$ from images of a calibration mask with known patterns at known positions. The image analysis and calibration algorithms described in this section have been implemented in Matlab$^\textrm{\textregistered}$. The process below is described for one camera only for the sake of readability, as it only have to be repeated for each camera independently in the case of a multi cameras system.

\begin{figure*}[h]
\centering
\includegraphics[width=.95\columnwidth]{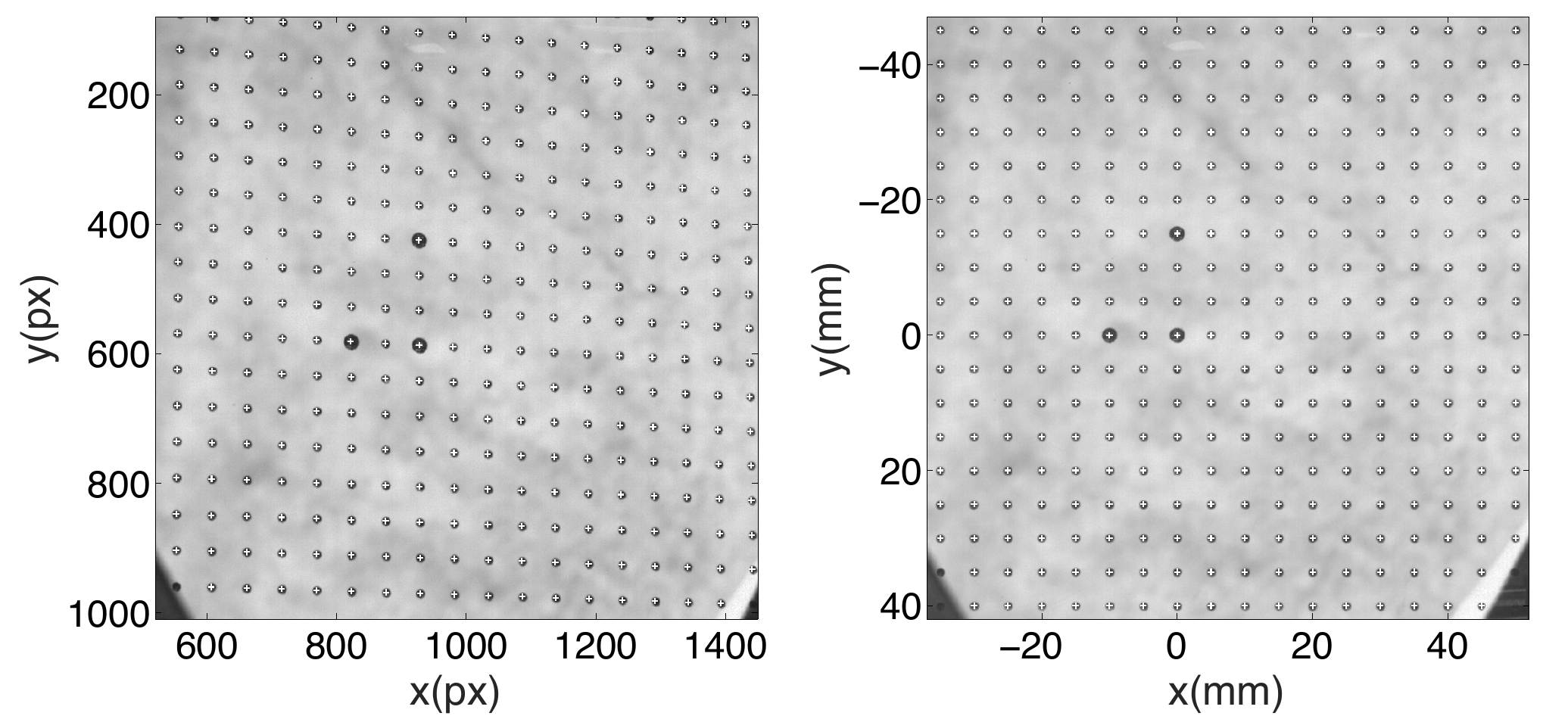}
\caption{a) Distorted raw image in pixel coordinates; the white crosses are the detected targets. b) Transformed image to the real world coordinates in mm.} 
\label{fig:px2mm}
\end{figure*}

We use a calibration mask, made of a grid of equally separated dots. Three points of the grid are bold marked in order to unambiguously define the $XOZ$ axes in \emph{real world} absolute coordinates. The calibration mask can be moved perpendicularly to its plane (along the $OZ$ axis) using a micrometric screw. Images of the calibration mask at $N_Z$ different known $Z$ positions are taken from both cameras: ${\cal{I}}_{j}$ is the calibration image when the plane is at the position $Z_j$ (with $j\in[1,N_Z]$). For the purpose of testing the quality of the new calibration method, up to $N_Z=13$ planes across a measurement volume have been taken (we will discuss the influence of the number of planes used to achieve the calibration in section \ref{sec:comp_tsai}). Figure~\ref{fig:px2mm}a shows an example of a typical calibration image of the mask. We then perform the following processing steps:

\begin{enumerate}

\item \textbf{Dot centers finding.} For each calibration image ${\cal{I}}_{j}$ we identify the centers of the dots, using standard center finding algorithms (in the present work the image of individual dots on the mask have {a diameter of the order of 10 pixels}, we therefore simply use the weighted center of mass as an accurate sub-pixel estimation of dots center). This gives a  set $(x_{j}^k,y_{j}^k)_{k\in[1;N_{j}]}$ of pixel coordinates of center of dots, where $N_{j}$ is the number of dots actually detected on each image ${\cal{I}}_{j}$ (we restate here the meaning of the indices: $Z_j$ represents the position of the calibration mask and $k$ is the index of detected dots). Absolute coordinates of the dots in \emph{real world} $(X_j^k,Y_j^k,Z_j^k)_{k\in[1;N_{j}]}$ are known (relatively to a given position). In what follows, lowercase coordinates represents \emph{pixel coordinates}, while uppercase coordinates represent \emph{absolute real world coordinates}. The detected center dots have been reported in Fig.~\ref{fig:px2mm}a. 

\item \textbf{2D Plane by plane transformations.} For each position $Z_j$ of the calibration mask we use the known 2D pixel coordinates $(x_{j}^k,y_{j}^k)_{k\in[1;N_{j}]}$ and the known 2D absolute coordinates $(X_{j}^k,Y_{j}^k)_{k\in[1;N_{j}]}$ to infer a spatial transformation ${\cal{T}}_{j}$ projecting 2D pixel coordinates onto 2D real world coordinates in the plane $XOY$ at position $Z_j$. The inverse transformation ${\cal{T}}^{-1}_{j}$ is also simultaneously inferred. ${\cal{T}}_{j}$ and ${\cal{T}}^{-1}_{j}$ allow to transform back and forth pixel coordinates into real world coordinates in a plane $XOY$ attributed at $Z=Z_j$:

\begin{align*}
{{\cal{T}}_{j}}:	&\;\;  	\textrm{pixel array of the  camera} 	& \longrightarrow 	&\;\; \textrm{real world}\\
			&\;\;	(x,y) 								& \longrightarrow 	&\;\; (X,Y,Z_j)\\
			&									&			&	\\
{{\cal{T}}_{j}}^{-1}: 	&\;\; \textrm{real world} 	& \longrightarrow & \;\;\textrm{pixel array of the camera} \\
				&\;\; (X,Y,Z_j)		& \longrightarrow & \;\;(x,y) 
\end{align*}

Different type of transformations can be inferred, from a simple linear projective transformation, to high order polynomial transformations if non-linear optical aberrations need to be corrected (main optical aberrations are properly captured by a 3rd order polynomial transformation). This is a standard planar calibration procedure, commonly used for instance in 2D PIV (and implemented in most commercial PIV softwares).  In practice, the direct and inverse transformations  can be efficiently estimated using the \texttt{cp2tform} function in Matlab$^{\textregistered}$. As an example, Figure~\ref{fig:px2mm}b shows the same image as in~\ref{fig:px2mm}a after a 3rd order polynomial transformation has been applied. The crosses reported on the image correspond to the actual absolute position of the center of dots expected for the undistorted image. An estimate of the accuracy of the 2D plane by plane transformation can be obtained from the distance, in pixel coordinates, between $(x_{j}^k,y_{j}^k)_{k\in[1;N_{j}]}$ and ${\cal{T}}^{-1}_{j}\left(X_{j}^k,Y_{j}^k,Z_j\right)_{k\in[1;N_{j}]}$. The maximum error for the image presented in Figure~\ref{fig:px2mm}a is less than 2 pixels.

\item \textbf{Building the pixel-line interpolant.} The last and key step of the present calibration method aims at building a pixel-line interpolant, ${\cal I}$, which directly connects pixels coordinates to a ray path. To achieve this, we define a grid of $N_{\cal{I}}$ interpolating points in pixel coordinates $(x_{l}^{\cal{I}},y_{l}^{\cal{I}})_{l\in[1,N_{\cal{I}}]}$ {for which the ray paths have to be computed}. 
{We chose to built the interpolant using all pixels of each camera, as this step is only done once, but it can also be done with a sub-sample of the array of pixels of the camera}. Then, we use the inverse transformations ${\cal{T}}^{-1}_{j}$ to project each point of this set back onto the real world planes {$(X,Y,Z_j)$} at each of the $N_Z$ positions $Z_j$. Each interpolating point $(x_{l}^{\cal{I}},y_{l}^{\cal{I}})$ is therefore associated to a set of $N_Z$ points in real world $(X_{l}^{\cal{I}},Y_{l}^{\cal{I}},Z_j)$. Conversely, these points in real world can be seen as a discrete sampling of the ray path which impacts the sensor of the camera at $(x_{l}^{\cal{I}},y_{l}^{\cal{I}})$. If we assume light propagates as a straight line, the $N_Z$ points $(X_{l}^{\cal{I}},Y_{l}^{\cal{I}},Z_j)$ are therefore supposed to be aligned. By a simple linear fit of these points, we can then directly relate each interpolating point $(x_{l}^{\cal{I}},y_{l}^{\cal{I}})$ in pixel arrays to a line $\Delta_{l}$ defined by one point $O_{\Delta_{l}}=(X^0_{l},Y^0_{l},Z^0_{l})$ and one vector $\vec{V}_{\Delta_{l}}=(Vx_{l},Vy_{l},Vz_{l})$ (hence 6 parameters in total). Finally we use these pixel-line correspondences from the $N_{\cal{I}}$ interpolation points to infer the interpolant ${\cal{I}}$, which allows to relate any pixel coordinate $(x,y)$ in the camera to the ray path $(O_\Delta,\vec{V}_\Delta$) corresponding to all possible positions of light sources producing a bright spot in $(x,y)$. In practice, the interpolant ${\cal{I}}$ is composed of 6 interpolants (one for each of the 6 parameters parameter defining the line of the ray path) estimated from the corresponding data defined in the 2 dimensional grid formed by the array of interpolating points $(x_{l}^{\cal{I}},y_{l}^{\cal{I}})_{l\in[1,N_{\cal{I}}]}$. 



\end{enumerate}

\subsection{Comparation with Tsai method}\label{sec:comp_tsai}
\begin{figure*}[h]
\centering
\includegraphics[width=.65\columnwidth]{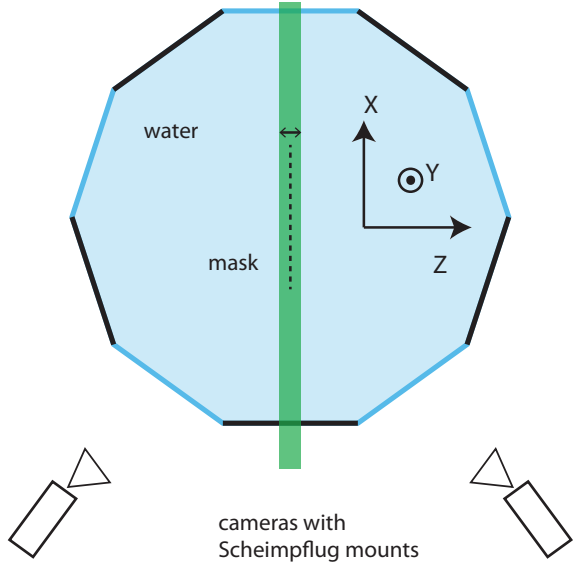}
\caption{Top view of the stereoscopic optical arrangement used for Particle Tracking in the LEM flow. Both camera use an objective with Scheimpflug mount so that all objects in a $10 \times 10 \times 1$ cm$^3$ region near the geometrical center is approximately in focus on the camera sensor. The calibration mask is placed parallel to the light sheet and moved in the $Z$ direction.} 
\label{fig:setup}
\end{figure*}

Since the publication of the calibration procedure proposed by Tsai \cite{bib:tsai1987}, a wide number of experimentalists have used this method to recover the optical characteristics of a camera necessary to reconstruct the 3D perspective of an object. Currently this method is widely used to calibrate the cameras for Lagrangian tracking techniques in fluid dynamics. In order to grasp the accuracy of the proposed camera calibration procedure, we compare it with the Tsai technique below.
{The stereoscopic optical arrangement is sketched on figure \ref{fig:setup}. It aims at performing Particle Tracking Velocimetry in a thick laser sheet near the geometrical center of a turbulent water flow created in an icosahedron (the LEM flow, further described in sec \ref{sec:setup}). Each camera objective, nearly perpendicular to its corresponding window, is mounted in the Scheimpflug configuration so that all particles present in the laser sheet are nearly in focus whatever their position in the field of view.} To perform both calibrations, we used a translucent plate mounted parallel to the laser sheet with dots size equal to 2 mm. These points were equally spaced by a distance of 5 mm in both directions and the thickness of the plate was approximately 0.2 mm. This plate was attached to a manual micro-metric traverse that was able to give displacements with a precision of the order of ten micrometers. For both methods, 13 images of the target were used, spaced 1 mm from each other along the $Z$ axis. The interpolant for the proposed method was built {considering a simple line} equation, as light is expected to propagate straight in the system used.

\begin{figure}[h]
\centering
\includegraphics[width=.45\columnwidth]{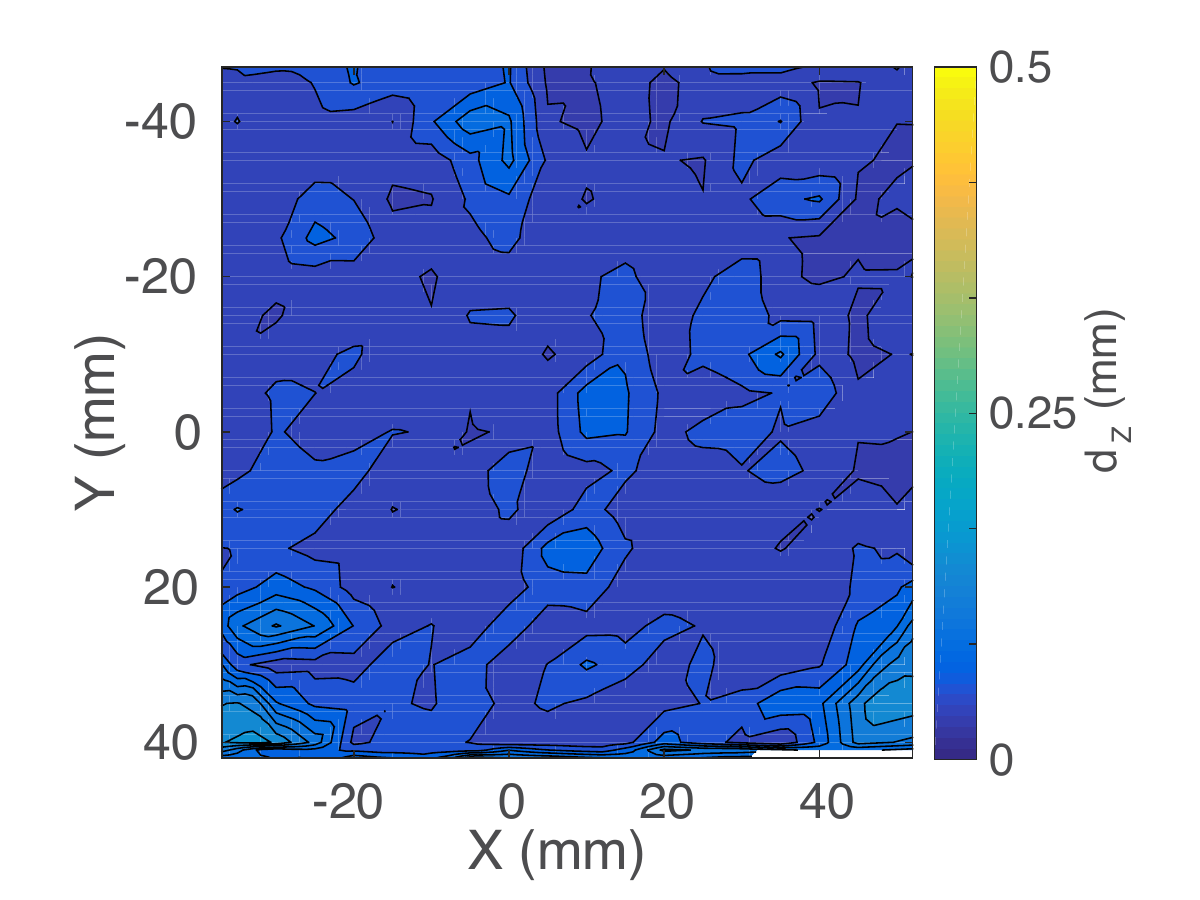}
\includegraphics[width=.48\columnwidth]{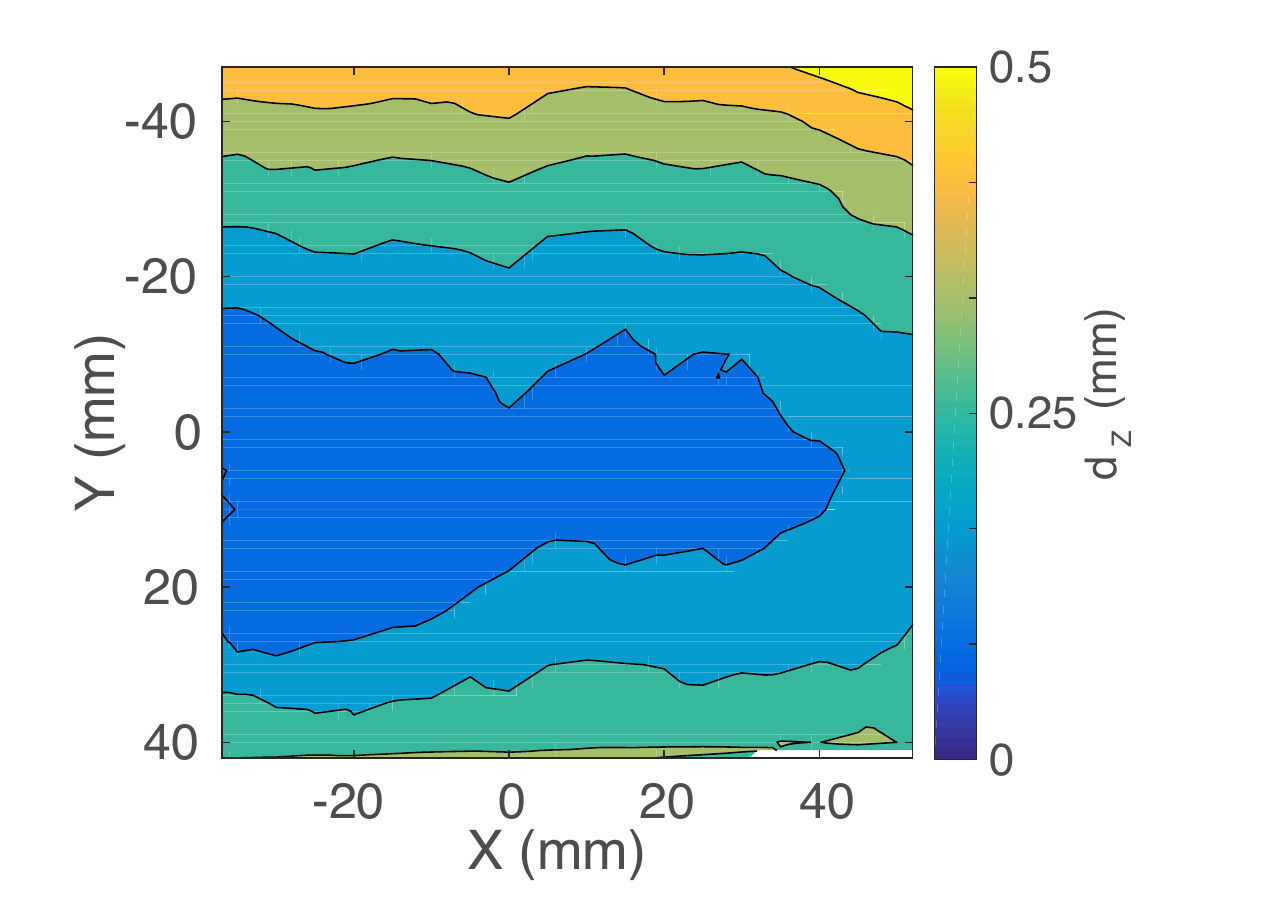}
\caption{Contour fields of the calibration error averaged over $Z$ (\textit{i.e.} over the 13 positions of the calibration plate), using the proposed calibration method (a) or Tsai model (b).} 
\label{comcalibzoom}
\end{figure}

On each plane of the 3D space where the target was imaged, we know exactly the 3D positions of the dots but also their 2D measured positions. Applying the calibration on the latter gives a series of positions that cannot match exactly the real coordinates because, whatever the method, {the parameters are obtained by solving an over-constrained linear system in the least-square sense.} We can then calculate the calibration error, \textit{i.e.} the absolute difference between the real coordinates and the transformed ones, for each plane, to evaluate the calibration accuracy. {Because we work with a stereoscopic imaging system, the mean distance between the dots measured and real positions for each dot $j$ on each plane $k$ is estimated after stereo matching their positions in real world coordinates using the two camera views. This distance can be estimated only along one direction (we note for example $d_X$ as the $X$-error) or as a sum over the three directions $d=\sqrt{ d_X^2+ d_Y^2+ d_Z^2}$}. Figure \ref{comcalibzoom} plots the latter averaged on the 13 planes used, for both the proposed method and the Tsai model.

 
Even if light propagates along straight lines in our system, we see that the accuracy of the proposed calibration is much better, as can be also appreciated with the numerical values given in Table \ref{tabla1}. The error is overall at least 300\% smaller (depending on which component is considered) and is reduced to barely half a pixel. It is also important to note that the error map obtained with the Tsai method (Fig. \ref{comcalibzoom}b) seems to display a large bias along the $Y$ coordinate (direction perpendicular to the plane that contains the cameras' optical axes) that could be due to the use of Scheimpflug mounts, which are typically not included in the Tsai calibration, and to the angle between the cameras and the test section windows. 
{This hypothesis was verified by comparing the two calibrations procedures in more conventional conditions, {\it i.e.} with sets of calibration images obtained without Scheimpflung mounts. In such cases, both methods proved to give similar results with very small error.}\\
{For the present optical arrangement and the new calibration method, we find the error in the $Y$ positioning is smallest. Indeed, due to the shape of the vessel (an icosahedron), the $y$ axis of the camera sensor is almost aligned with the $Y$ direction {so that this coordinate is fully redundant between the views,} while the $x$ axis of the camera sensor forms an angle $\alpha \simeq \pi/3$ with the $X$ direction so that the precision on $X$ positioning is lower. This directly impacts the precision on the $Z$ positioning whose error is almost equal to the $X$ positioning error.}


\begin{table}[h]
 \caption{\label{tabla1} \small Absolute deviation from the expected position of the targets averaged over space.}
\begin{center}
 \begin{tabular}{l c c c c}
   & $d_X$ ($\mu$m) & $d_Y $ ($\mu$m) & $d_Z$ ($\mu$m) & $d$ ($\mu$m) \\
 \hline
 Proposed calibration & 32.7 & 12.6 & 39.2 & 59 \\
 Tsai model & 121 & 171.1 & 112.7 & 266.6 \\
  
 \end{tabular}
\end{center}

 \end{table}

\subsection{Discussion}
Up to 13 planes were used to build the operator that yields the camera calibration. While two planes are the minimum required for the method, a larger number of planes imaged provide better results in calibration accuracy. In the case study presented here as an example, the major sources of optical distortion were the classic ones: Scheimpflug mounts, imperfect lenses and non perpendicular interfaces. The use of 7 planes provided an optimal trade-off of high accuracy and simplicity in the calibration procedure. The average error using 7 planes is only 2$\%$ larger than that using all 13 planes, while using 3 planes in the calibration results in an error that is $10\%$ larger. If dealing with a more complex system, for instance one with an optical index gradient, increasing the number of planes used in the calibration could improve the results, as it allows the calibration operator to more accurately capture the curvature of the light rays.  

The new proposed calibration method described here has several advantages that made it worth implementing in a multi-camera particle imaging experimental setting. First, it requires no models or assumptions about the properties of the optical path followed by the light from the scattering element in the flow to the camera sensor. The method computes an operator, an interpolator of high polynomial order,  that determines the equation for propagation of light in space. This ray line equation is fully determined by the physical location of the calibration dots (light scatterers) distributed in known positions in space. Secondly, this method is turnkey for any existing optical system (where a calibration method like Tsai's was previously applied). The planes imaged in an optical experimental setting (to which the traditional Tsai calibration was formerly applied) can be used with the new calibration method proposed here. The implementation of the new method proposed here was initially developed in Matlab$^\textrm{\textregistered}$, which could be done using only existing routines. Implementation in an open source language, such as Python, would be trivial. Finally, it is important to note that using the calibration operator, being a polynomial interpolant, to compute the physical position of points in regions of the images that were not covered by the calibration plane images, can lead to large errors. This is of course also true for model-based distortion corrections, as the constants in the models are only valid in the region imaged in the calibration process.


\section{A multi-time-step noise reduction method for measuring velocity statistics from Particle {Tracking} Velocimetry.}
\label{sec:stat}

\subsection{Particle location measurement noise background}

Any experimental measurement is necessarily contaminated to a certain extent by noise. This is particularly critical for measurements of velocity fields where differentiation is frequently needed. The use of particles, tracers for flow velocity measurements or inertial particles for particle dynamics measurements, introduces noise both from the light scattering and imaging physics, and from the tracking/correlation mathematical processing of those images\cite{adrian1991particle,bib:Toschi:ar2009}. 

Generally described, in a PTV experiment the particle positions are determined, including measurement noise, by detecting the centroid of each region of the camera sensor that records light intensity above a certain threshold in contiguous pixels. While the uncertainty in the detection of the centroid as the true location of the particle center does not represent a significant problem in the computation of particle location statistics, the noise in the particle location becomes a severe limitation when using these measurements to calculate velocity or acceleration (first and second derivatives, respectively). The need to filter the particle positions prior to any differentiation has long being recognized {\cite{Voth:jfm2002,mordant:2003,Volk:epl2007,berg:2009}}. Multiple filter parameters, for example filter length, have to be chosen or adjusted to minimize the propagation and amplification of noise in the process of taking the derivative of the measurements. How the derived quantities depend on the filtering can be studied to provide confidence in the data, even leading to information on the expected value of some moments of the measurement distribution (the rms value of particle acceleration in \cite{mordant:2003,Volk:epl2007}). Filtering, however, generally removes information, erasing events (changes in the particle trajectories) with time scales shorter than the filter length. 

PIV experiments treat noise differently than PTV. Individual particle positions are not determined, but rather a correlation of the position of several particles as an ensemble is used to compute flow displacement (and therefore velocity) at a certain scale. The impact of particle location noise in the raw images is thus reduced by the convolution (integral) algorithm compared to {filtering and }differentiation in PTV. 

The second contribution in this manuscript presents a {simple} method that uses individual tracking of particles, \`a la PTV, to compute Eulerian statistics of the velocity (similar to those obtained from PIV), including resolution of the smallest scales of the flow. 
It uses a multi-time-step approach that takes advantage of the double frame capability of PIV-enabled cameras or the fast image acquisition of high speed cameras (or both in PIV-enabled high speed cameras). These capabilities allow for several sequences of image pairs to be taken with PIV-enabled cameras at increasing time intervals between images in the pair, or to skip different number of frames in a high speed image sequence. Thus, several reconstructions of the velocity field in the flow are obtained from the tracking of particle positions between two sets of images spaced by different intervals of time ($\Delta~t_{frames 1-2} = n \times \tau$, where {$\tau$} is the smallest characteristic time used in imaging the flow, and n is an increasing integer - 1, 2, 3, ...). The velocity fields obtained in this manner are then used to compute different realizations of the velocity statistics, and the {absence} of correlation between the particle location noise used to remove the dominant source of noise from the statistic of interest, for example the velocity structure function. This opens the door for more accurate calculations of turbulent quantities such as dissipation.

\subsection{Multi-time-step method for velocity field noise correction. Theoretical Basis}
\label{Theory}

{In PTV experiments, one usually obtains an estimate of the velocity from the Lagrangian displacement field of particles which have been identified in consecutive frames separated by an interval $\Delta t$. We define it at the position $\mathbf{X}(t)$ and note it $\Delta \mathbf{X}= \mathbf{X}(t+\Delta t)-\mathbf{X}(t)$. Because the measured particle positions contain noise $\mathbf{b}$ (coming from centroid detection, imaging artifacts such as astigmatism, ...) added to the real positions $\widetilde{\mathbf{X}}$, statistics computed using a naive definition of the velocity $\Delta \mathbf{X}/\Delta t$ are biased. This section discusses how to extract noiseless estimates of the mean and variance velocity field and its extension to the estimate of a noiseless second order structure function. The method uses common properties of the noise $\mathbf{b}$ : it is frequency independent (white noise), has zero temporal mean, and is decorrelated from the real signal, such that:}
\begin{equation}
   \begin{array}{ccccc}
    \left\langle \mathbf{b}(t)\mathbf{b}(t+\Delta t) \right\rangle_t &=& \left\langle \mathbf{b}^2 \right\rangle_t \delta(\Delta t) & &\\\\ 
    \left\langle \mathbf{b}.\Delta\widetilde{\mathbf{X}} \right\rangle_t &=& \left\langle \mathbf{b} \right\rangle_t. \left\langle \Delta \widetilde{\mathbf{X}} \right\rangle_t &=  0, \nonumber\\ \\ 
    \end{array}
\end{equation}

\noindent where $\left\langle . \right\rangle_t$ is a time (or ensemble) average. The only reasons to have noise correlated to the signal are systematic stereo-matching or particle detection errors. We allow for the noise to be correlated in space, which can easily happen due to lighting inhomogeneities or sensor sensitivity gradients, for instance. The displacement field is, then, $\Delta \mathbf{X}$=$\Delta \widetilde{\mathbf{X}}$ +$\Delta \mathbf{b}$. While the first moment of the displacement field can be free of noise if $\left\langle \mathbf{b} \right\rangle_t=0$, its second moment can be expressed as:


\begin{equation}
   \begin{array}{ccl}
    \left\langle (\Delta \mathbf{X})^2\right\rangle_t &=& \left\langle \widetilde{\mathbf{v}}^2\right\rangle_t \Delta t^2 + 2\left\langle \mathbf{b}^2\right\rangle_t + 2\left\langle \widetilde{\mathbf{a}}.\widetilde{\mathbf{v}}\right\rangle_t \Delta t^3 +o(\Delta t^3),\\\\ 
    \end{array}
\label{eq:dx2}
\end{equation}
using a $2^{nd}$ order Taylor expansion $\Delta \widetilde{\mathbf{X}}=\widetilde{\mathbf{v}}\Delta t+\frac{1}{2}\widetilde{\mathbf{a}}\Delta t^2+o(\Delta t^3)$. If we bin the Lagrangian displacement measurements by their position in space, and subtract the mean displacement in each bin (free of noise as shown above), we obtain an Eulerian field of fluctuating displacement that is of interest to characterize the flow structures in time or space. {As seen in equation \ref{eq:dx2}, $\langle (\Delta \mathbf{X})^2 \rangle_t$ has a first contribution proportional to the second order moment of the real fluctuating velocity $\left\langle \widetilde{\mathbf{v}}^2\right\rangle_t$, a second one equal to twice the noise variance, plus a correction ($\Delta t^3$) that depends on the cross correlation of the fluctuating velocity and acceleration.}

The new method to remove the noise from the velocity statistics consists, simply, in calculating {$\left\langle \Delta \mathbf{X}\right\rangle_t$ and} $\left\langle (\Delta \mathbf{X})^2\right\rangle_t$ for multiple experiments where images of the particles in the flow are collected at increasing values of $\Delta t$. {At a first step, these statistics can be computed on a grid, $X_\text{grid}$, to obtain the Eulerian values $\left\langle \Delta \mathbf{X}\right\rangle_t(X_\text{grid})$ and $\left\langle (\Delta \mathbf{X})^2\right\rangle_t(X_\text{grid})$ (at each location, $X_\text{grid}$, in the fluid domain onto which the Lagrangian displacements have been binned).}
{Because the mean displacement field is noiseless, it is found to be proportional to $\Delta t$ so that the mean velocity is $\left\langle \mathbf{\widetilde{v}}\right\rangle_t (X_\text{grid}) = \left\langle \Delta \mathbf{X}\right\rangle_t(X_\text{grid})/\Delta t$. This is well satisfied in experiments and serves as a test that the noise has zero mean.\\ 
The evolution of $\left\langle (\Delta \mathbf{X})^2\right\rangle_t(X_\text{grid})$ with $\Delta t$ is then fitted at each point of the grid by first order polynomials of the form $\alpha \Delta t^2 + \beta$, the leading coefficients being directly the values of the velocity variance $\left\langle \mathbf{\widetilde{v}}^2\right\rangle_t(X_\text{grid})$. The reason not to take the third order correction into account relies on properties of turbulent flows for which $\left\langle \widetilde{\mathbf{a}}.\widetilde{\mathbf{v}}\right\rangle_t$ is well approximated by the dissipation rate $\varepsilon$. From dimensional analysis, one then gets an estimate of the ratio $\left\langle \widetilde{\mathbf{v}}^2\right\rangle_t / \left\langle \widetilde{\mathbf{a}}.\widetilde{\mathbf{v}}\right\rangle_t \tau_\eta \sim Re_\lambda$, where $\tau_\eta = \sqrt{\nu/\varepsilon}$ is the dissipative time and $Re_\lambda$ is the Reynolds number at the Taylor length-scale. Taking time increments $\Delta t$ smaller than the dissipative time then ensures that the displacement field is well approximated by a parabola in high Reynolds number flows.}

\begin{figure}[h]
\centering
\includegraphics[width=.5\columnwidth]{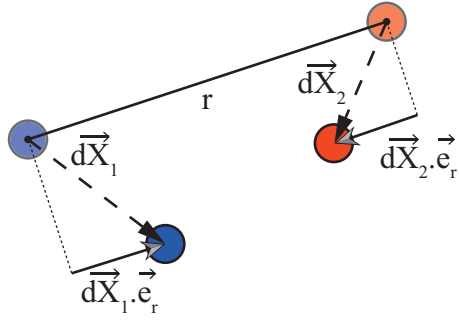}
\caption{Sketch of the calculation of the displacement difference between two particles separated by a distance $r$. To compute the longitudinal second order structure function, the component of the particle displacement computed from each particle position at two different times is projected onto the line that connects the particle centers (at the first time, $t$). The projection line between particles is shown with solid black line. The particle displacement is shown with dashed arrow. The displacement component projected onto the inter-particle line is shown as a solid arrow (with grey point). The perpendicular component of the displacement, not used for the computation of the longitudinal structure function, is shown as a dashed line. The particle positions for this purpose (at the first time, $t$) are shown with faint colors, while the particle positions at the later time ($t+dt$) are shown with solid colors.} 
\label{fig:str_fct_method2}
\end{figure}

This method can be extended to higher order moments of the displacement field, as well as be used to recover increment statistics, more particularly the {longitudinal} second order structure function of the real velocity $\widetilde{S_2}=\left\langle [(\mathbf{\widetilde{v}(X+r)}-\mathbf{\widetilde{v}(X)}).\mathbf{e_r}]^2\right\rangle$, where {$\mathbf{e_r} =\mathbf{r}/r$ and}  $\left\langle . \right\rangle$ is a time and space average. One just has to estimate:
\begin{equation}
   \begin{array}{ccl}
    \left\langle [(\Delta \mathbf{X(X+r)}-\Delta\mathbf{X(X)}).\mathbf{e_r}]^2\right\rangle &=& \left\langle [(\Delta \mathbf{b(X+r)}-\Delta\mathbf{b(X)}).\mathbf{e_r}]^2\right\rangle\\
    & & +\left\langle [(\mathbf{\widetilde{v}(X+r)}-\mathbf{\widetilde{v}(X)}).\mathbf{e_r}]^2\right\rangle \Delta t^2\\
    & & +\left\langle [(\mathbf{\widetilde{v}(X+r)}-\mathbf{\widetilde{v}(X)}).\mathbf{e_r}][(\mathbf{\widetilde{a}(X+r)}-\mathbf{\widetilde{a}(X)}).\mathbf{e_r}]\right\rangle \Delta t^3+o(\Delta t^3)
    \end{array}
\label{eq:S2}
\end{equation}
where $\left\langle [(\Delta \mathbf{b(X+r)}-\Delta\mathbf{b(X)}).\mathbf{e_r}]^2\right\rangle$ {does not depend on $\Delta t$, but only on $\mathbf{r}$ if the noise statistics are homogeneous in space.}
{From equation \ref{eq:S2}, one can see that a noiseless estimate of the structure function is obtained by fitting the evolution of $\left\langle [(\Delta \mathbf{X(X+r)}-\Delta\mathbf{X(X)}).\mathbf{e_r}]^2\right\rangle$ with $\Delta t$, at each separation $| \mathbf{r}|$, using again a first order polynomial $c_1 \Delta t^2 + c_2$ whose leading coefficient is $\widetilde{S_2}(r)$.}
The structure functions computed with this method are for the velocity component aligned with the line between particle positions, $\mathbf{r}$, as sketched in Fig.~\ref{fig:str_fct_method2}. Note that the structure function computation does not require the conversion of the displacement field to Eulerian coordinates, but rather to bin {only} the inter-particle distance $| \mathbf{r}|$. This means that measuring structure functions is always possible at arbitrarily small separations $| \mathbf{r}|$, without any requirements on the Eulerian spatial binning (that depends on the number of particle pairs in the flow). Contrary to that, this method requires only a statistical convergence in the number of particles at a certain range of inter-particle distance (a number that goes with $N^2$). This represents a significant advantage of this method over computation of structure functions from PIV measurements, where small particle separations, $| \mathbf{r}|$, requires very small interrogation windows with the corresponding increase in measurement noise.


The second order moment of the velocity fluctuations and second order structure function are presented in this theoretical derivation as examples of what the Taylor expansion of the statistical moments, combined with data collected at different $\Delta t$ can achieve. Higher order moments (for both the velocity fluctuations and the structure functions) can be easily computed with this method. The only difference with the values presented here are that those higher moments are contaminated with the residual noise left behind by the computation of the lower order moments. 

\subsection{Practical Implementation of the Method.}
\subsubsection{Flow Set-up}
The experimental set-up (drawn in figure \ref{fig:setup}) consists of a large icosahedral vessel filled with deionized water. The flow is forced by 12 propellers driven at the same frequency $f_{prop}$ by servomotors. This excitation produces homogeneous and isotropic turbulence with zero mean velocity in the center of the chamber\cite{bib:lionel,bib:zimmermann}. Polystyrene particles with a diameter of 225$\mu$m ($\simeq 5 \eta$) and an average density of $\rho \sim 1.09$ g/cm$^3$ were used as tracers (Table \ref{tabla3}). The integral length scale of the flow $L_\text{int}=u'^3/\varepsilon\simeq 4$ cm and the values of the Kolmogorov length scales and time scales $\eta=(\nu^3/\varepsilon)^{1/4}$ and $\tau_\eta=(\nu/\varepsilon)^{1/2}$ (with $\nu=1\times 10^{-6}$ the kinematic viscosity of water at 20$^\circ$C) are computed using the velocity fluctuation rms, $u'$, and the turbulent kinetic energy dissipation, $\varepsilon$, quantities both obtained from second order structure function computed with the new method proposed here.

\begin{table}[h]
 \caption{\label{tabla3} Flow parameters for different propeller frequencies $f_{prop}$. $\varepsilon$: energy dissipation rate per unit mass; $\eta$: Kolmogorov length scale; $\tau_\eta$: Kolmogorov time scale; $R_\lambda=\sqrt{15{u'}^4/\varepsilon\nu}$: Taylor-microscale Reynolds number, $u'$ being the velocity fluctuation intensity {and $\nu$ the kinematic viscosity of the fluid}.}
 \begin{center}
\begin{tabular}{c c c c c c}
  	$f_{prop}$ & $u'$ & $\varepsilon$ & $\eta$ & $\tau_\eta$  & $R_\lambda $ \\
  	$Hz$ & cm/s & $m^2 s^{-3}$ & $\mu m$ & $ms$ & \\
 	\hline
 	8  & 15 & 0.09 & 56 & 3.2 & 291 \\
 	10 & 17 & 0.17 & 49 & 2.4 & 339 \\
 	12 & 24 & 0.33 & 42 & 1.7 & 388 \\
 \end{tabular}
  \end{center}
 \end{table}

\subsubsection{Particle Imaging Set-up}\label{sec:setup}
Two CMOS camera with a resolution of $2048 \times 1088$ pixels were used in a stereoscopic arrangement. Images were collected in double frame mode, each camera collected two frames separated by a time $\Delta t$, and the rate of acquisition of pairs was {125 Hz}. We used a Nd:YAG laser (wave length of 532 nm) to produce a laser sheet approximatively 1 cm thick. Measurements were  obtained in a volume of $10 \times 10 \times 1$ cm$^3$. We restrict the  analysis to a smaller volume of $70 \times 70 \times 2$ mm$^3$ to ensure that the flows is homogeneous and isotropic. For each experiment, approximatively 10 000 pairs of image sets (each set providing the 3D position of several hundred particles in the flow) were collected to ensure statistical convergence.

The new particle displacement analysis method described in this paper requires to repeat the experiments (keeping the flow conditions, such as the Reynolds number, constant) with different values of the camera and illumination settings so the time-step between consecutive images, $\Delta t$, varies. Alternatively, a very fast acquisition/illumination rate using high speed camera and KHz pulsed lasers allows to collect a single experimental image sequence and then take a variable $\Delta t$ in the analysis by skipping an increasing number of images in the sequence. In this implementation, we collected five different experiments, for each flow Reynolds number studied, with five different values of $\Delta t$. The maximum value was approximately $20\%$ of $\tau_\eta$,  and the other four values were equal to some fraction of the maximum. Values are given in Table \ref{tabla4}.

\begin{table}[h]
 \caption{\label{tabla4} Values of the time-steps $\Delta t_i$ for the different {propeller frequencies} explored.}
  \begin{center}
\begin{tabular}{c c c c c c c}
  	$f_{prop}$ & $\tau_\eta$ & $\Delta t_1/\tau_\eta$ & $\Delta t_2/\tau_\eta$ & $\Delta t_3/\tau_\eta$ & $\Delta t_4/\tau_\eta$ & $\Delta t_5/\tau_\eta$ \\
  	$Hz$ & $ms$ & & & & & \\
 	\hline
 	8 & 3.2 &  0.03 & 0.06 & 0.10 & 0.12 & 0.15 \\
 	10 & 2.4 & 0.04 & .08 & .12 & .17 & .21 \\
 	12 & 1.7 & 0.05 & 0.09 & .12 & .17 & .23 \\
 \end{tabular}
  \end{center}
 \end{table}

\subsection{Results}\label{sec:res_dt}
\begin{figure}[h]
\includegraphics[width=.49\columnwidth]{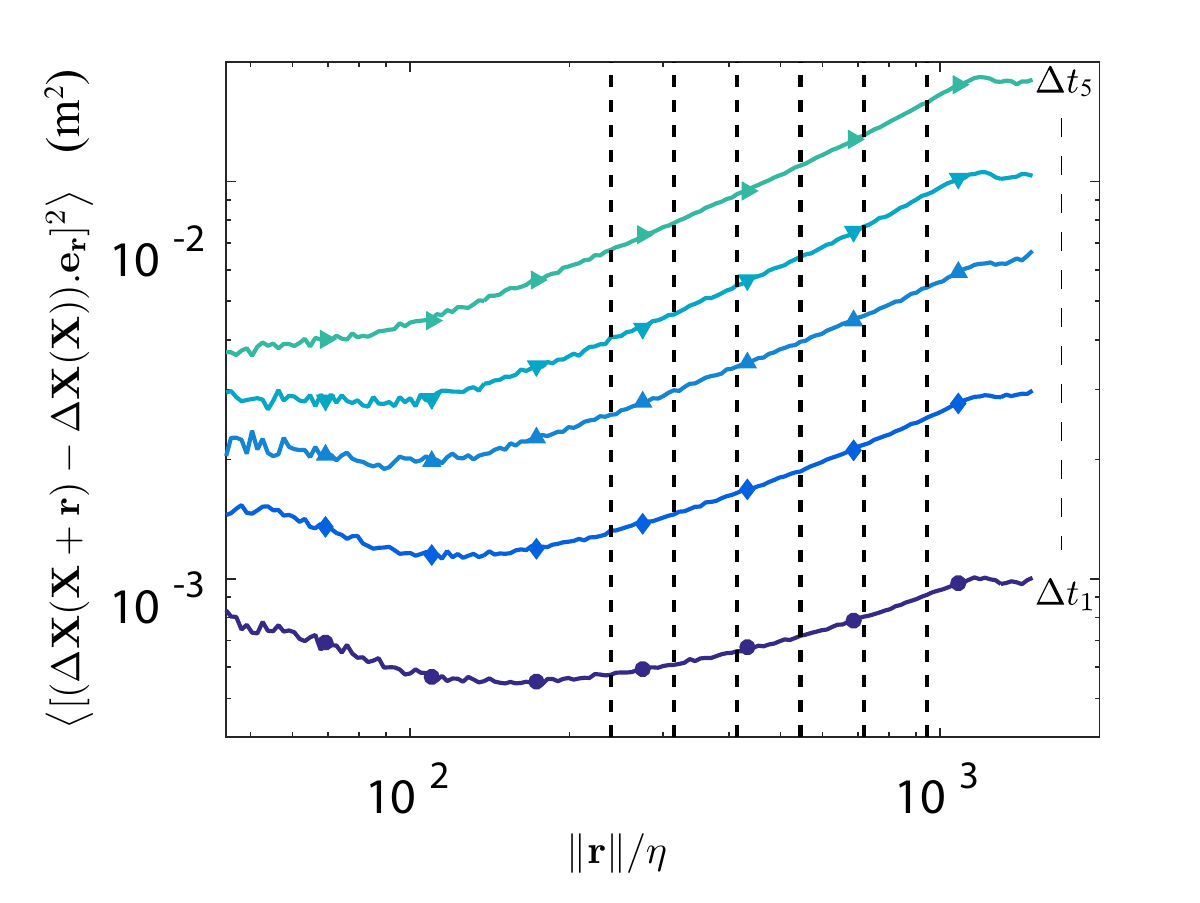}
\includegraphics[width=.49\columnwidth]{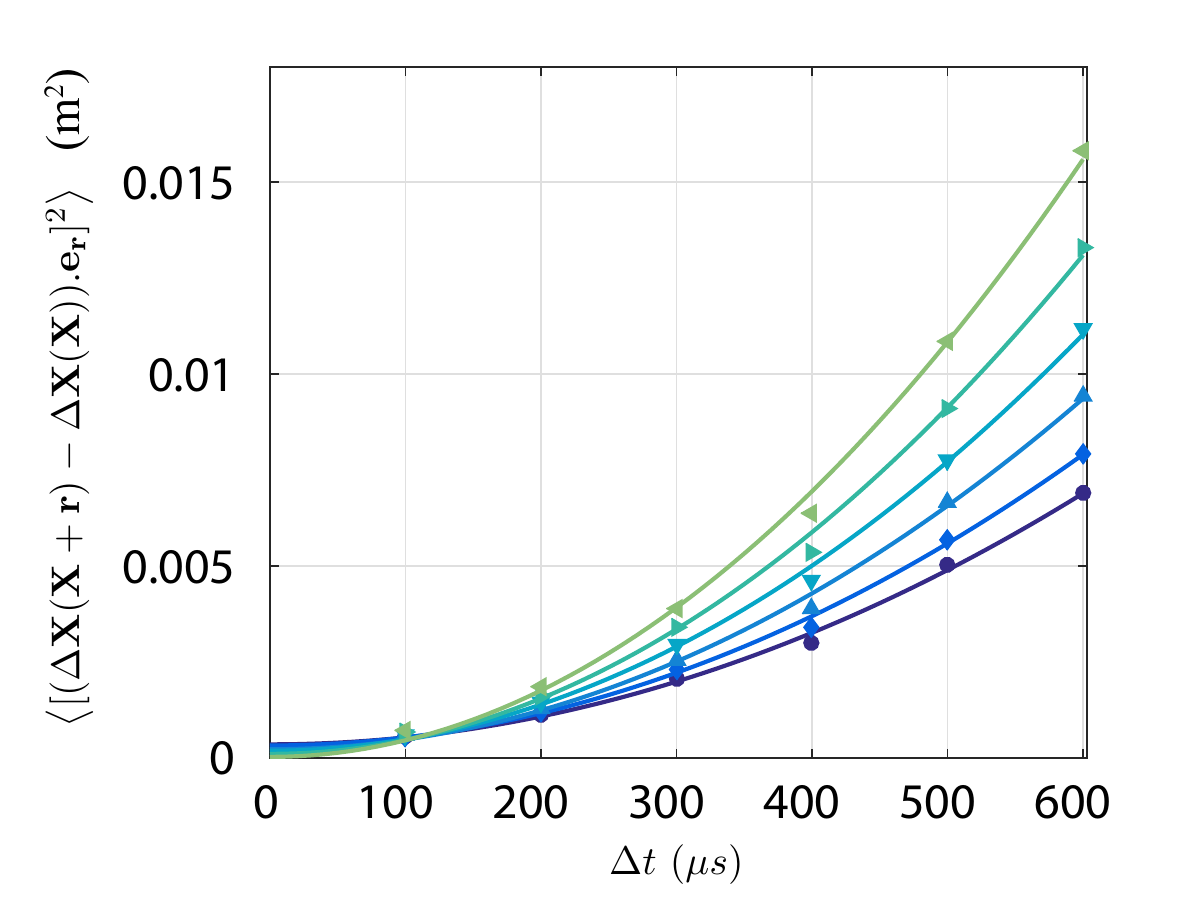}
\caption{a) Longitudinal second order structure functions of the raw displacement field $\Delta \mathbf{X}$ against the separation $|\mathbf{r}|$ normalized by the Kolmogorov length scale $\eta$ for different values of $\Delta t_i$ at $Re=3.1 \times 10^5$ (see Table \ref{tabla4}). b) Same quantities but plotted at a given separation $|\mathbf{r}|$, as a function of the inter-frame time-step value $\Delta t$. The six different symbols and lines in (b) stand for different values of $|\mathbf{r}|$, indicated by the vertical dashed lines on (a); ascending order on (b) is for increasing value of separation (left to right on a). The lines are fits of the form $c_1\Delta t^2+c_2$.}
\label{dtstf}
\end{figure}

\begin{figure}[h]
\centering
\includegraphics[width=.49\columnwidth]{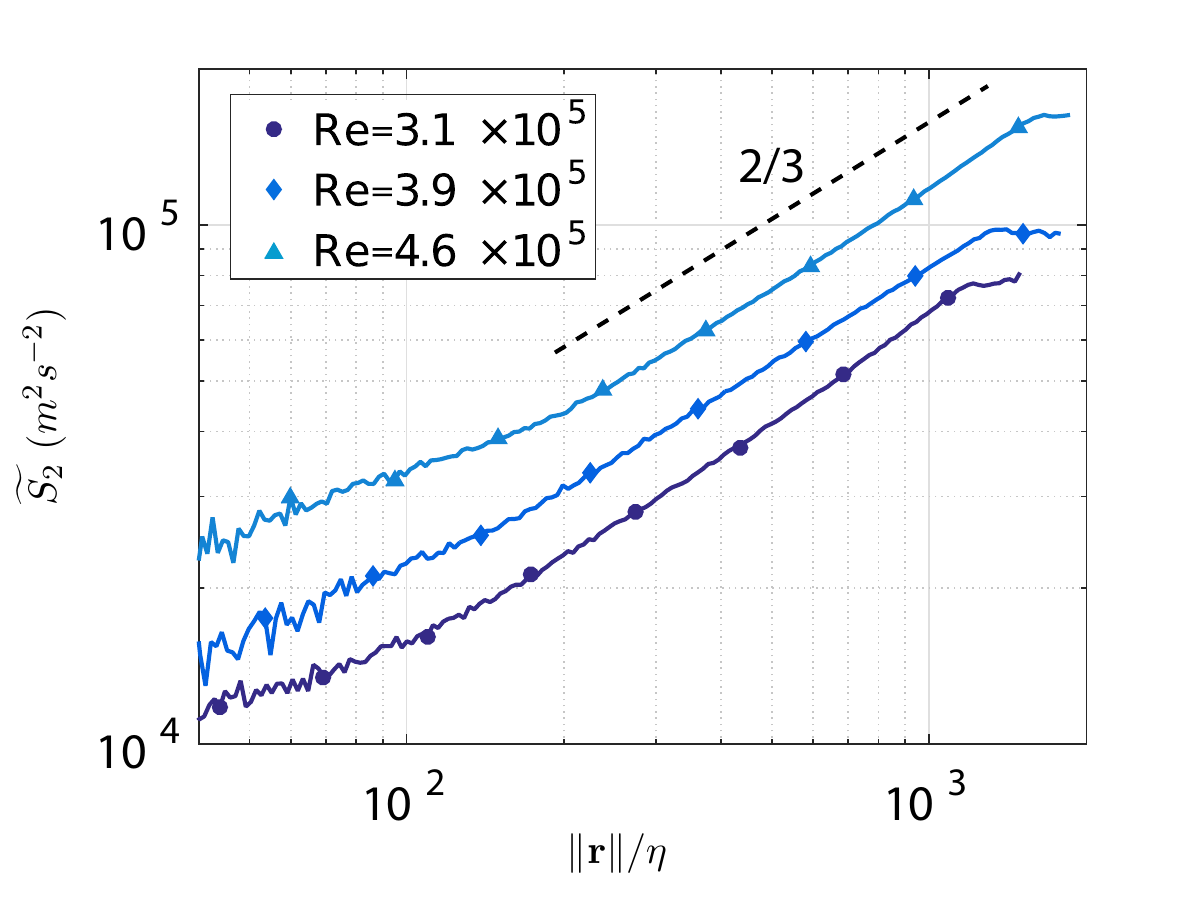}
\includegraphics[width=.49\columnwidth]{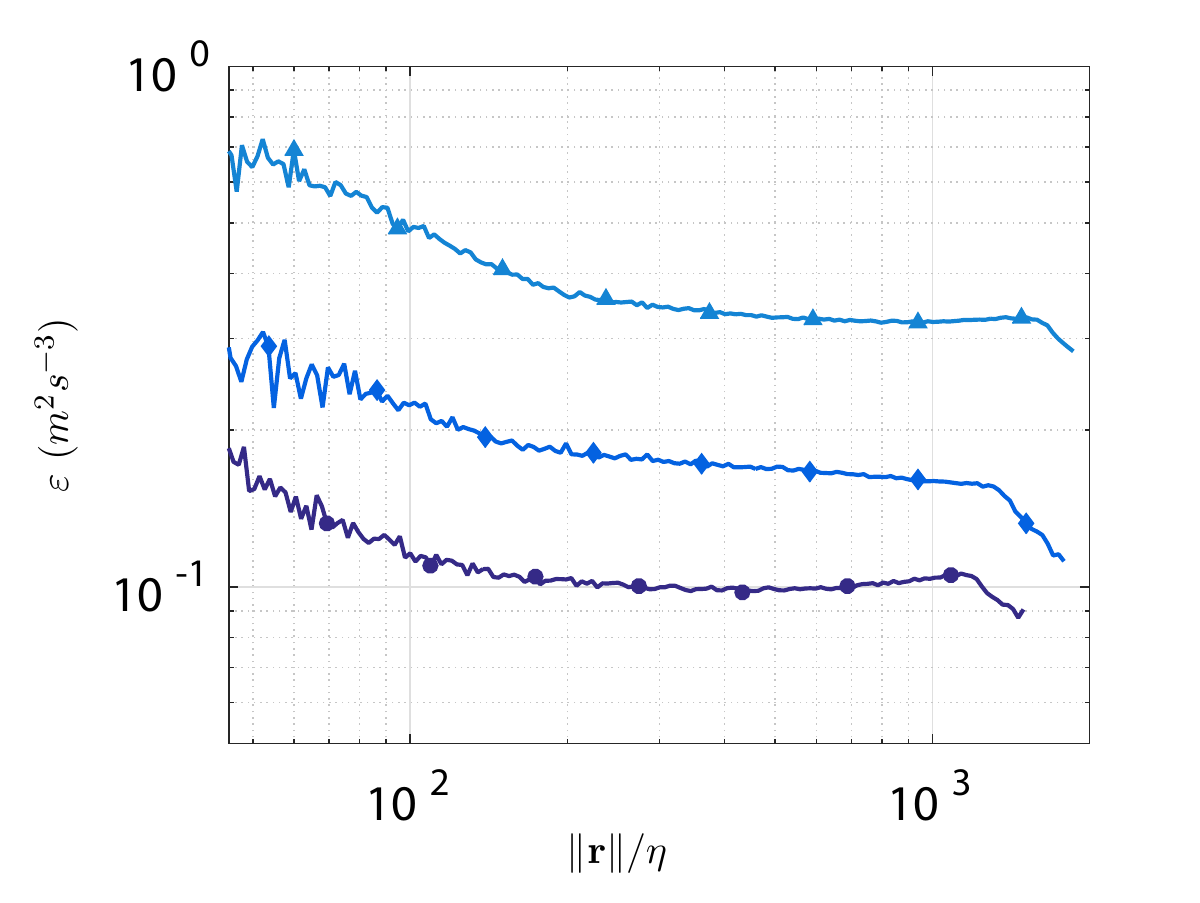}
\caption{a) Second order structure functions of the velocity extracted with the proposed method for different $Re$ values. The black dashed line corresponds to a power law of exponent $2/3$. b) Energy dissipation rate estimated as $\varepsilon_r=\widetilde{S_2}^{3/2}/|\mathbf{r}|$.}
\label{strfun2}
\end{figure}

Figure \ref{dtstf}a shows the longitudinal second order structure functions of the displacement field $\Delta \mathbf{X}$. We observe strong changes of the shape of the functions that come from how the noise affects the signal for a given value of the time-step $\Delta t$. The different values of the structure functions for different $\Delta t$ follows $\Delta t^2$ (equation \ref{eq:S2}) as the dependence of the function with $\Delta t$ is independent of the noise. We plot this evolution in Figure \ref{dtstf}b for 6 values of the separation $|\mathbf{r}|$. Fitting this data with a quadratic a function of the form $c_1\Delta t^2+c_2$ yields good agreement, with positive values of the coefficient $c_2$ for all experiments at different Reynolds numbers. This positive constant is proportional to the variance of the noise (as shown in section~\ref{Theory}. The value of the coefficient $c_1$ in the quadratic fit with time, for each value of the particle separation $|\mathbf{r}|$ is exactly the second order function of the velocity, with the noise removed, as plotted in Figure \ref{strfun2}a for three different values of the flow Reynolds number. Note the presence of the inertial range, denoted by the 2/3 slope in the log-log plot, over approximatively one decade. The $2/3$ power law displayed by $\widetilde{S_2}$ corresponds well with Kolmogorov's prediction for homogeneous isotropic turbulence $\widetilde{S_2}\sim \varepsilon^{2/3}|\mathbf{r}|^{2/3}$ \cite{kolmogorov1941local}. This type of turbulence variables extracted from velocity measurements would be subject to a significant level of uncertainty and inaccuracy, if the noise from the raw particle position was not removed with the new process described in this paper (as can be seen in Figure \ref{dtstf}(a)). 

Figure \ref{strfun2}b shows the estimation of the  dissipation rate of turbulent kinetic energy, $\varepsilon_r=\widetilde{S_2}^{3/2}/|\mathbf{r}|$ for three different Reynolds numbers studied in this experimental implementation of the denoising method. The plateaux found in the decade of separation values shown confirm the presence of the inertial range and their values correspond to the ensemble average of the local dissipation rate. The values of the dissipation in the flat region where it does not depend on particle separation (averaged over the inertial range) were the source of the values of $\varepsilon$ in Table \ref{tabla3}. We have not shown here maps of the fluctuating velocity obtained by the proposed method (for the sake of briefness).
The spatial average of the fluctuating velocity is the value of $u'$ in Table \ref{tabla3}. Those values compared well with those in \cite{bib:lionel}, obtained by 2D3C PIV, confirming the accuracy of the method. In fact, the values of $u'$ and $\varepsilon$ are slightly lower than those obtained by PIV. This discrepancy can be explained, qualitatively, based on the physics of the measurements and the effect of the noise on these metrics when it is not eliminated from the displacement measurements. Equation \ref{eq:dx2} shows that the velocity measured by PIV is equal to the real velocity plus the noise variance (potentially convoluted to each other). The structure function (and hence $\varepsilon$) is also subject to this erroneous increase in the value due to the creep-in of the noise into the statistical value computation. Equation \ref{eq:S2} shows that the term $\left\langle [(\Delta \mathbf{b(X+r)}-\Delta\mathbf{b(X)}).\mathbf{e_r}]^2\right\rangle$ will increase the value $\varepsilon$ due to noise. To determine the importance of this term, it has to be expanded into $4\left\langle \mathbf{b}^2\right\rangle(1-C_b(|\mathbf{r}|))$, where $C_b(|\mathbf{r}|)$ is the noise spatial correlation, bounded between (-1, 1). Regardless of the value of  $C_b$, it will increase erroneously the value of the structure function yielding a higher value of $\varepsilon$ and because the value of $C_b$ depends on spatial separation, it will not raise it uniformly for all values of $|\mathbf{r}|$), changing the slope of the structure function with separation, thus making the value of $\varepsilon$ noisier.

\subsection{Discussion}\label{sec:disc_dt}
The comparison of the flow statistics with a previous 2D3C PIV study allows for the validation of the proposed method. In fact, the measurements show better results, with no need to tune arbitrary filtering parameters to remove noise (\sout{\textit{i.e.}} the interrogation window size {for instance}). The only parameters that have to be chosen for the new method proposed here are the different values of $\Delta t$ that are accessible for a given flow and camera/illumination available, the form of the fit function and finally the binning in space to compute the Eulerian average and fluctuating velocities, and in separation distance to compute the structure function.

The values of $\Delta t$ are subject to two limitations. They have to be high enough so that particles move more than the measurement error. They need to be small enough that the large displacement associated with high value of $\Delta t$ does not interfere with the ability of the particle tracking algorithm to identify individual particles \cite{ouellette2006experimental}. 
{As mentioned in the previous section, a maximum value of $\Delta t \lesssim \tau_\eta$ ensures that the third order correction in equations \ref{eq:dx2} remains small as one has $u'^2 / \left\langle \widetilde{\mathbf{a}}.\widetilde{\mathbf{v}}\right\rangle_t \tau_\eta \sim Re_\lambda$. This was verified in the present experimental set-up and we found this correction to be negligible compared to the second order term. This was also the case for the structure function provided the separation lies in the inertial range $|\mathbf{r}| \gg \eta$. In such cases,} the best agreement between fit functions and the data overall was found when using a quadratic function of $\Delta t$. As for the number of time-step values needed, the value of $\varepsilon$ when using only the 3 larger values of $\Delta t$ was only $5\%$ lower than using all six experiments. Using only the lowest value and largest values of $\Delta t$ allowed for a simple calculation of $\varepsilon$ that was only $2\%$ higher than with the full experimental dataset.


The displacement vector field obtained from Particle Tracking in this multi-time-step method is computed in a Lagrangian frame of reference. To compute the values of $\left\langle (\Delta \mathbf{X})^2\right\rangle_t$ against $\Delta t$, the displacement field must be binned into a spatial grid, converting it to an Eulerian frame of reference. 
Although the number of particles per image, or Eulerian grid cell, is relatively small in this PTV images, the velocity is estimated independently for each particle pair. Thus, the statistical convergence in the method is reached relatively soon (with the need for a very large number of image pairs). The computation of the structure functions highlights this advantage even more. {As pointed out above in section~\ref{Theory}, the structure function should in principle be computed to arbitrarily small separation between particles using the method. However great care has to be taken in doing so because: i) of the difficulty to achieve statistical convergence in finding particles with small separations. ii) the second and third order terms in equation \ref{eq:S2} will become of the same order of magnitude as the separation enters the dissipation range $|\mathbf{r}| \sim 10 \eta$. These may be the reasons why we always observe an increase of the structure functions at small separations.}


This new method requires each experiment and/or analysis to be repeated several time, for different values of $\Delta t$, while keeping everything constant in the experiment. Access to noise-free turbulence variables, enabled by advances in imaging, laser and computational technology, seems like a reasonable trade-off for the additional effort. A major advantage of the proposed method is that the signal is never differentiated to obtain velocity statistics. If one is only interested in inertial range or larger scale velocity statistics, the use of larger values of $\Delta t$ (with an eventual reduction of the seeding density to allow for successful  tracking) would still make the computation of the structure functions possible.


\section{Conclusions}
In this article, we have presented two methods to improve the accuracy of flow or particle velocity measurements. The first contribution consists on a calibration method that does not require an optical model of the camera, lenses, test section windows, Scheimpflug, etc. It should be trivial to implement the technique with both the calibration target image acquisition and target points spatial location being identical to what is currently standard in the field. Programming the calibration algorithm and the operator to convert pixel locations to physical locations, with minimal errors, could be done in any language and any computer available to fluid dynamics experimentalists. We prove that the new method is at least equal and frequently more accurate that the commonly-used the Tsai model as it can be used in a wider range of optical configurations. As experimental set-ups get more complicated with more optical elements and more light refraction opportunities, the new method should prove simpler to implement, more effective and much more accurate than the model-based Tsai. 

The second contribution presented here, is a new method proposed to remove measurement noise from the calculation of velocity statistics, without filtering. It has been tested to compute the value of the energy dissipation rate through the second order structure function of the velocity. An experiment imaging tracer particles in a turbulent flow via stereoscopic particle tracking was used to demonstrate the concept. The combination of both methods allowed for an estimation of the flow statistics in good agreement with a previous 2D3C PIV study, and we believe with better accuracy.

\begin{acknowledgements}
{The authors would like to acknowledge the financial support of the EUHIT project "European High performance Infrastructures in Turbulence" funded by the European Commission under Grant agreement no: 312778. This research was conducted during A. Aliseda's stay as an invited professor at LEGI Grenoble and ENS de Lyon. Financial support from Labex TEC 21 (Investissements dÕAvenir - grant agreement n¡ANR-11-LABX-0030) and ENS de Lyon is acknowledged. This work is supported by the European French research programs ANR-12-BS09-0011, and Projet Emergent PALSE/2013/26.}
\end{acknowledgements}

\end{document}